\documentclass{aastex61}

\received{\today}
\submitjournal{ApJ}

\shorttitle{A Detached Protostellar Disk around a Very Low-Luminosity Object}
\shortauthors{Tokuda et al.}

\begin{document}

\title{A Detached Protostellar Disk around a $\sim$0.2$M_{\odot}$ protostar in a Possible Site of a Multiple Star Formation in a Dynamical Environment in Taurus}

\correspondingauthor{Kazuki Tokuda}
\email{tokuda@p.s.osakafu-u.ac.jp}

\author[0000-0002-2062-1600]{Kazuki Tokuda}
\affiliation{Department of Physical Science, Graduate School of Science, Osaka Prefecture University, 1-1 Gakuen-cho, Naka-ku, Sakai, Osaka 599-8531, Japan}
\affiliation{Chile Observatory, National Astronomical Observatory of Japan, National Institutes of Natural Science, 2-21-1 Osawa, Mitaka, Tokyo 181-8588, Japan}

\author{Toshikazu Onishi}
\affiliation{Department of Physical Science, Graduate School of Science, Osaka Prefecture University, 1-1 Gakuen-cho, Naka-ku, Sakai, Osaka 599-8531, Japan}

\author{Kazuya Saigo}
\affiliation{Chile Observatory, National Astronomical Observatory of Japan, National Institutes of Natural Science, 2-21-1 Osawa, Mitaka, Tokyo 181-8588, Japan}

\author{Takashi Hosokawa}
\affiliation{Department of Physics, Kyoto University, Sakyo-ku, Kyoto 606-8502, Japan}

\author{Tomoaki Matsumoto}
\affiliation{Faculty of Sustainability Studies, Hosei University, Fujimi, Chiyoda-ku, Tokyo 102-8160, Japan}

\author{Shu-ichiro Inutsuka}
\affiliation{Department of Physics, Nagoya University, Chikusa-ku, Nagoya 464-8602, Japan}

\author{Masahiro N. Machida}
\affiliation{Department of Earth and Planetary Sciences, Faculty of Sciences, Kyushu University, Nishi-ku, Fukuoka 819-0395, Japan}

\author{Kengo Tomida}
\affiliation{Department of Earth and Space Science, Osaka University, Toyonaka, Osaka 560-0043, Japan}

\author{Masanobu Kunitomo}
\affiliation{Department of Physics, Nagoya University, Chikusa-ku, Nagoya 464-8602, Japan}

\author{Akiko Kawamura}
\affiliation{Chile Observatory, National Astronomical Observatory of Japan, National Institutes of Natural Science, 2-21-1 Osawa, Mitaka, Tokyo 181-8588, Japan}

\author{Yasuo Fukui}
\affiliation{Department of Physics, Nagoya University, Chikusa-ku, Nagoya 464-8602, Japan}

\author{Kengo Tachihara}
\affiliation{Department of Physics, Nagoya University, Chikusa-ku, Nagoya 464-8602, Japan}



\begin{abstract}

We report ALMA observations in 0.87 mm continuum and $^{12}$CO ($J$ = 3--2) toward a very low-luminosity ($<$0.1 $L_{\odot}$) protostar, which is deeply embedded in one of the densest core MC27/L1521F, in Taurus with an indication of multiple star formation in a highly dynamical environment. The beam size corresponds to $\sim$20 AU, and we have clearly detected blueshifted/redshifted gas in $^{12}$CO associated with the protostar. 
The spatial/velocity distributions of the gas show there is a rotating disk with a size scale of $\sim$10 AU, a disk mass of $\sim$10$^{-4}$ $M_{\odot}$ and a central stellar mass of $\sim$0.2 $M_{\odot}$.
The observed disk seems to be detached from the surrounding dense gas, although it is still embedded at the center of the core whose density is $\sim$10$^{6}$ cm$^{-3}$. 
The current low-outflow activity and the very low luminosity indicate that the mass accretion rate onto the protostar is extremely low in spite of a very early stage of star formation. We may be witnessing the final stage of the formation of $\sim$0.2 $M_{\odot}$ protostar.
However, we cannot explain the observed low luminosity with the standard pre-main-sequence evolutionary track unless we assume cold accretion with an extremely small initial radius of the protostar ($\sim$0.65 $R_\odot$).
These facts may challenge our current understanding of the low mass star formation, in particular the mass accretion process onto the protostar and the circumstellar disk.

\end{abstract}

\keywords{stars: formation --- circumstellar matter  --- stars: low-mass --- stars: protostars --- stars: individual (L1521F-IRS) --- ISM: kinematics and dynamics}



\section{Introduction} \label{sec:intro}
 Investigation of accretion activities onto new-born stars is a vital subject to understand origins of the stellar initial mass function (IMF), which is a fundamental issue of star formation. Despite of a large number of efforts, mechanisms to determine stellar masses are not clearly understood. In isolated low-mass star formation in dense molecular cloud cores, some early studies suggested that outflows from protostars eventually terminates gas accretion onto protostars \citep[e.g.,][]{Arce06,Machida13}. In cluster formation regime, numerical simulations suggested that dynamical interactions among formed protostars truncate protostellar disks with a size scale of a few tens au and terminate the mass accretion \citep[e.g.,][]{Bonnell03}. \\
\ To challenge such problems observationally, measuring masses of evolving young protostars, which has been used to trace the evolutionary stage, is important. The velocity/spatial distribution of Keplerian disks is powerful tool to measure the central protostellar mass. Recent interferometric observations, especially with ALMA, have revealed Keplerian disks around the Class 0 protostars \citep[e.g.,][]{Tobin12,Ohashi14,Aso17,Yen17}, although such disks are not expected to be observed when a protostar is at a very young stage \citep[e.g.,][]{Yen15} or when disk formation is suppressed by magnetic braking \citep[e.g.,][]{Allen03,Mellon08,Yen15}. 
Molecular line observations toward YSOs having Keplerian disks also show active outflows with a size scale of $\gtrsim$1000 AU and/or inflow motions, indicating that significant accretion activities onto the protostars and/or disks are continuing. These results are qualitatively consistent with the general picture of the evolution from a dense core to a protostar \citep[e.g.,][]{Shu87,Inutsuka12}. These observational studies are still limited to relatively bright ($\gtrsim$1 $L_{\odot}$) protostar. High-sensitivity infrared observatories, such as $Spitzer$, have identified much fainter ($L$ $<$ 0.1 $L_{\odot}$) protostars, the so-called $``$very low-luminosity Objects (VeLLOs)$"$ \citep[e.g.,][]{Young04,Dunham06,Lee09}. Some are deeply embedded in dense cores, which had been classified as $``$starless cores,$"$ until $Spitzer$ discovered VeLLOs therein.
Because the expected brightness in the standard model of star formation with an accretion rate of $\sim$2 $\times$ 10$^{-6}$ $M_{\odot}$\,yr$^{-1}$ exceeds 1 $L_{\odot}$ unless the protostellar mass is extremely small, it is difficult to explain the faint luminosity of the VeLLOs. Recent theoretical studies have attempted to explain the faintness of the VeLLOs by reconsidering the stellar evolution in the accretion phase and the following pre-main-sequence stages. For instance, \cite{Vorobyov17} argue that the majority of the VeLLOs should be in the Class I phase, where protostars have already grown in mass via accretion, if the accreting gas provides very low entropy to the protostar \citep[called $``$cold accretion,$"$ e.g.,][]{Hosokawa11}. \\
\ MC27 (a.k.a. L1521F) \citep[e.g.,][]{Mizuno94,Onishi96,Onishi98,Onishi99,Onishi02,Codella97} is one of the densest condensations in low-mass, star-forming regions with a peak density of $\sim$10$^{6}$ cm$^{-3}$ at a single-dish resolution ($\sim$20\arcsec), and it contains a VeLLO (L1521F-IRS) at the center of the core with the luminosity of $<$0.07 $L_{\odot}$ \citep{Bourke06,Terebey09} whose spectral energy distribution is consistent with a Class 0/I object. 
\cite{Tokuda14} (Paper I) and \cite{Tokuda16} (Paper II) recently observed this condensation with ALMA in Cycle 0/1 and presented the detailed gas/dust distributions with the size scales continuously from a few tens to $\sim$10,000 au by combining the ALMA (12 m + 7m array) and the single-dish observations. The ALMA observations revealed that the system might be a possible site of multiple star formation because a few starless high-density ($\sim$10$^6$--10$^7$ cm$^{-3}$) condensations (MMS-2,3) were found within a region of a several hundred au around the known protostar (Paper I, II). 
A mm-dust source (MMS-1), which is associated with the protostar, has not been spatially resolved even at the resolution of $\sim$0\farcs74 $\times$ 0\farcs34 and confirmed no gas association, indicating that the protostar does not have extended gas envelope with the size scale of more than a few tens au (Paper I, II). However, the compact outflow with the size of a few hundred au emitted from the source (Paper I, II) implies that there is a disk-like structure around the protostar. 
We also found highly complex gas envelopes, such as arc-like structures, 
indicating that the initial condition of the low-mass star formation is highly dynamical \citep[Paper I, see also numerical simulations by ][]{Matsumoto15}. Although the ALMA observations provided us the significant information that we have not seen with the previous instruments, further investigations were needed to understand the overall picture of this complex system. In particular, the evolutionally status of the very low-luminosity protostar is still unclear, possibly due to insufficient spatial resolutions.\\

\section{Observations} \label{sec:Obs}
We carried out ALMA Cycle 3 Band 7 (0.87 mm) continuum and molecular line observations toward MC27/L1521F with a center position of ($\alpha_{J2000.0}$, $\delta_{J2000.0}$) = (4$^{\rm h}$28$^{\rm m}$38\fs96, +26\arcdeg51\arcsec35\farcs0) with the ALMA 12 m array. The observations were carried out in September 2016. There were four spectral windows and the correlator was set to have a band width of 1875 MHz. Two of them have a frequency resolution of 0.9765 MHz (1920 channels), corresponding to the velocity resolution of $\sim$0.85 km\,s$^{-1}$ with the frequency domain mode (FDM). The rest was set to the time domain mode (TDM) with 128 channels. The aggregate bandwidth is thus 7.5 GHz. The observed frequencies with the FDM include some molecular lines, $^{12}$CO ($J$ = 3--2), H$^{13}$CO$^{+}$ ($J$ = 4--3), and C$^{17}$O ($J$ = 3--2). The UV range is 15-2800 $k\lambda$. The calibrations of the complex gains and the flux were carried out through observations of a quasar, J0510+1800, and the phase calibrations were performed using J0433+2905. The data were processed with the CASA (Common Astronomy Software Application) package \citep{McMullin07} version 4.7.2. We used the briggs weighting with the robust parameter of 0.5. The synthesized beam is $\sim$0\farcs18 $\times$ 0\farcs1 = 25 au $\times$ 14 au at a distance of Taurus \citep[$\sim$140 pc,][]{Elias78}. The (1$\sigma$) rms of molecular line emission, $^{12}$CO ($J$ = 3--2) and the 0.87 mm continuum are $\sim$1.6 mJy\,beam$^{-1}$ at a velocity resolution of $\sim$0.85 km\,s$^{-1}$ and $\sim$35 $\mu$Jy\,beam$^{-1}$, respectively. 

\section{Results} \label{sec:results}

\subsection{Gas/dust distributions toward MMS-1\label{subsec:MMS-1_obs}}
\begin{figure}[htbp]
\plotone{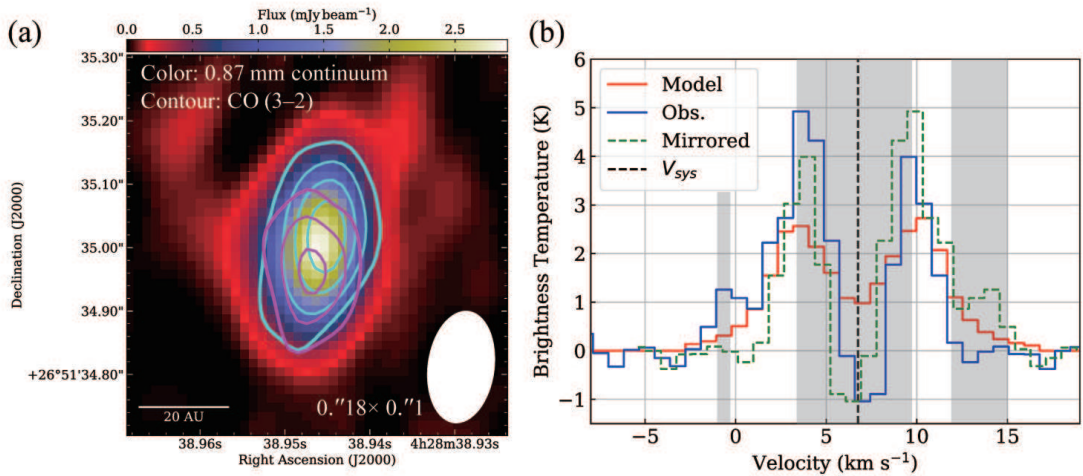}
\caption{Dust/gas distributions of a very low-luminosity protostar, MMS-1(L1521F-IRS) with ALMA. (a) Color-scale image shows the distribution of 0.87 mm continuum. Cyan and magenta contours show blueshifted ($-$0.65--4.45 km\,s$^{-1}$) and redshifted (8.7--12.1 km\,s$^{-1}$) components of $^{12}$CO ($J$ = 3--2), respectively. The lowest contour level and subsequent step are 0.014 Jy\,beam$^{-1}$\,km\,s$^{-1}$. The angular resolution is given by the ellipses in the lower right corner, 0\farcs18 $\times$ 0\farcs10. (b) Blue profile shows averaged $^{12}$CO ($J$ = 3--2) spectra over the region inside 5$\sigma$ detection in the 0.87 mm continuum. Red profile shows same one but for the simulated Keplerian disk model with $M{*}$ = 0.18 $M_{\odot}$ (see text). Green dashed line shows the observed profile mirrored about the systemic velocity, 6.75 km\,s$^{-1}$. Black dashed line denote the systemic velocity of 6.75 km\,s$^{-1}$. Dark hatched areas mark the velocity ranges in which there are discrepancies in intensity between the observation and the model (see the text for the possible explanation).
\label{fig:MMS-1}}
\end{figure}

\begin{figure}[htbp]
\plotone{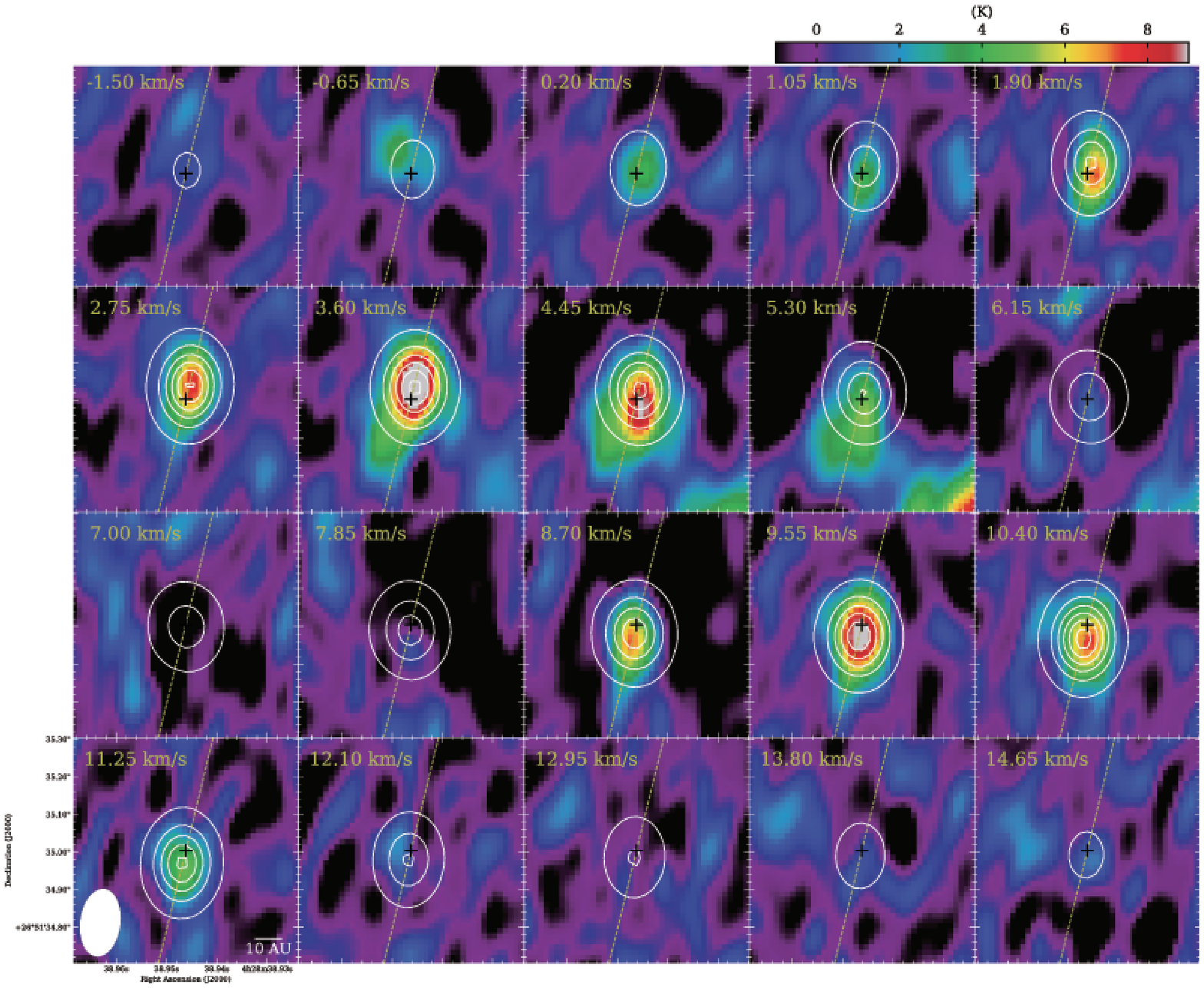}
\caption{Velocity-channel maps of the $^{12}$CO ($J$ = 3--2) emission toward MMS-1 with ALMA and those of a simulated Keplerien model (see the text). Color scales show the intensity maps of the ALMA observations with the velocity width for each map of 0.85 km\,s$^{-1}$. White contours in each panel represent the simulated Keplerian model. The lowest contour and subsequent steps are 0.5 K and 1.5 K, respectively. The central velocities are given in the upper left corner of each panel. The angular resolution of the $^{12}$CO ($J$ = 3--2) is given by the white ellipse in the  lower left corners of the bottom left panels. Black crosses in each panel represent position of the 0.87 mm continuum peak, ($\alpha_{J2000.0}$, $\delta_{J2000.0}$) = (4$^{\rm h}$28$^{\rm m}$38\fs9462, +26\arcdeg51\arcsec35\farcs004). The position angle of the disk, $-$17\arcdeg, is shown in yellow dashed lines.
 \label{fig:chanmap}}
\end{figure}

Figure \ref{fig:MMS-1} (a) shows the intensity maps of the 0.87 mm continuum and $^{12}$CO ($J$ = 3--2) observations toward MMS-1 at the angular resolution of $\sim$0\farcs18 $\times$ 0\farcs10. MMS-1 is marginally resolved in the dust continuum observation. We deconvolved the source in the image plane by using the point spread function (PSF) of the synthesized beam. The deconvolved size of MMS-1 is (91.1$\pm$10.1 mas)$\times$(55.8$\pm$4.2 mas), corresponding to (12.9$\pm$1.4 au)$\times$(7.8$\pm$0.6 au). The peak flux and the total flux are 2.8 mJy\,beam$^{-1}$ and 3.6 mJy, respectively, corresponding to the peak H$_2$ column density and the total mass of 3.1 $\times$ 10$^{23}$ cm$^{-2}$ and 8.3 $\times$ 10$^{-5}$ $M_{\odot}$ with the assumptions of the optically thin emission, the uniform dust temperature of 100 K, and the dust opacity per unit (gas + dust) mass column density,
$\kappa_{345 GHz}$ = 1.2 $\times$ 10$^{-2}$ cm$^2$\,g$^{-1}$ (\citealt{Hildebrand83,Oss94,Kauffmann08}; Paper I), which means that gas-to-dust ratio is assumed to be 100. We have clearly detected the blueshifted and redshifted emission of the $^{12}$CO ($J$ = 3--2) at MMS-1 (Figure \ref{fig:MMS-1} (a,b)) for the first time at the present high-angular resolution. The channel map of the $^{12}$CO ($J$ = 3--2) is shown in Figure \ref{fig:chanmap}. A position angle of the line passing through the emission peaks in each velocity channel is $-$17\arcdeg, which is measured counterclockwise relative to the north celestial pole. The angle is nearly perpendicular to that of the bipolar nebula seen in the $Spitzer$ observations \citep{Bourke06,Terebey09}. These gas/dust distributions strongly indicate that the rotating disk-like envelope exists around the very low-luminosity protostar, and the disk orientation is consistent with the direction of the outflow and the bipolar nebula (Paper I). The observed gas components around the systemic velocity of this core (6.15--7.85 km\,s) are not seen due to the optical thickness of the line.\\
\ As indicated by Paper I, II, our present observations also show that MMS-1 does not have extended high-density envelopes connecting to the disk except for the slightly elongated structures seen toward northwest and east directions in the 0.87 mm continuum (Figure \ref{fig:MMS-1} (a)). It is to be noted that we cannot detect high-density tracers, H$^{13}$CO$^{+}$ ($J$ = 3--2) and C$^{17}$O ($J$ = 3--2), toward MMS-1 even with the higher angular resolution compared with Paper II. These results indicate that the disk may be detached from the surrounding environment although it is located at the center of the dense core whose density is $\sim$10$^6$ cm$^{-3}$. 
We will discuss possible reasons to realize such an detached disk in Section \ref{subsec:detached}.

\subsection{Physical quantities of disk of MMS-1\label{subsec:MMS-1_model}}
\begin{figure}[htbp]
\plotone{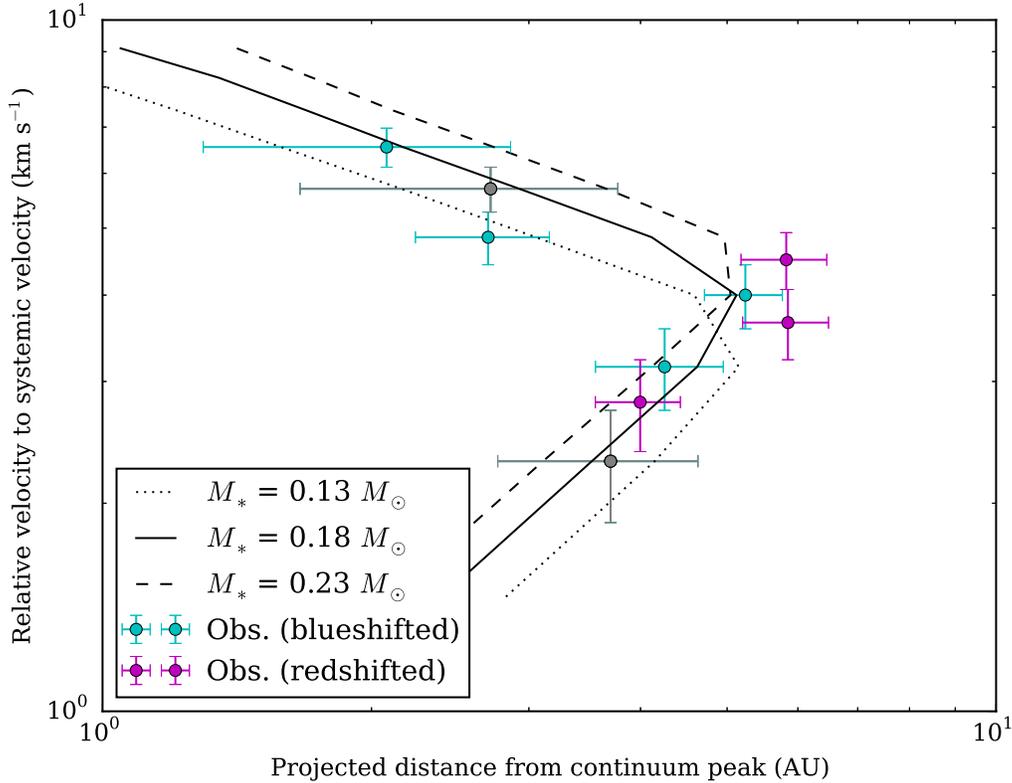}
\caption{Distance versus velocity profile of $^{12}$CO ($J$ = 3--2) emission toward MMS-1 in both the observations (circles) and the simulated models (lines). Cyan color and magenta color indicate blueshifted and redshifted emissions in the observations, respectively. Y-axis shows the relative velocity to the systemic velocity of 6.75 km\,s$^{-1}$. X-axis shows the projected distance from the 0.87 mm continuum peak to the $^{12}$CO ($J$ = 3--2) peak position derived by Gaussian fitting at each velocity channel (Figure \ref{fig:chanmap}). The uncertainty in the observed data is derived from the channel width and the positional uncertainty of the Gaussian fitting to the $^{12}$CO ($J$ = 3--2) emission. The black line present expected positions derived from the Keplerian disk model with the $M_{*} = 0.18\,M_{\odot}$.  The black dotted and dashed lines show those of the Keplerian models with $M_{*}$ = 0.13 $M_{\odot}$ and 0.23 $M_{\odot}$, respectively. Two peak positions shown in grayscale are located at the opposite side of the expected location, possibly due to the low signal-to-noise ratio and/or the contamination of the envelope emission. We do not take into account these points in the comparison with the model.
\label{fig:plot}}\end{figure}

\begin{figure}[htbp]
\epsscale{0.5}
\plotone{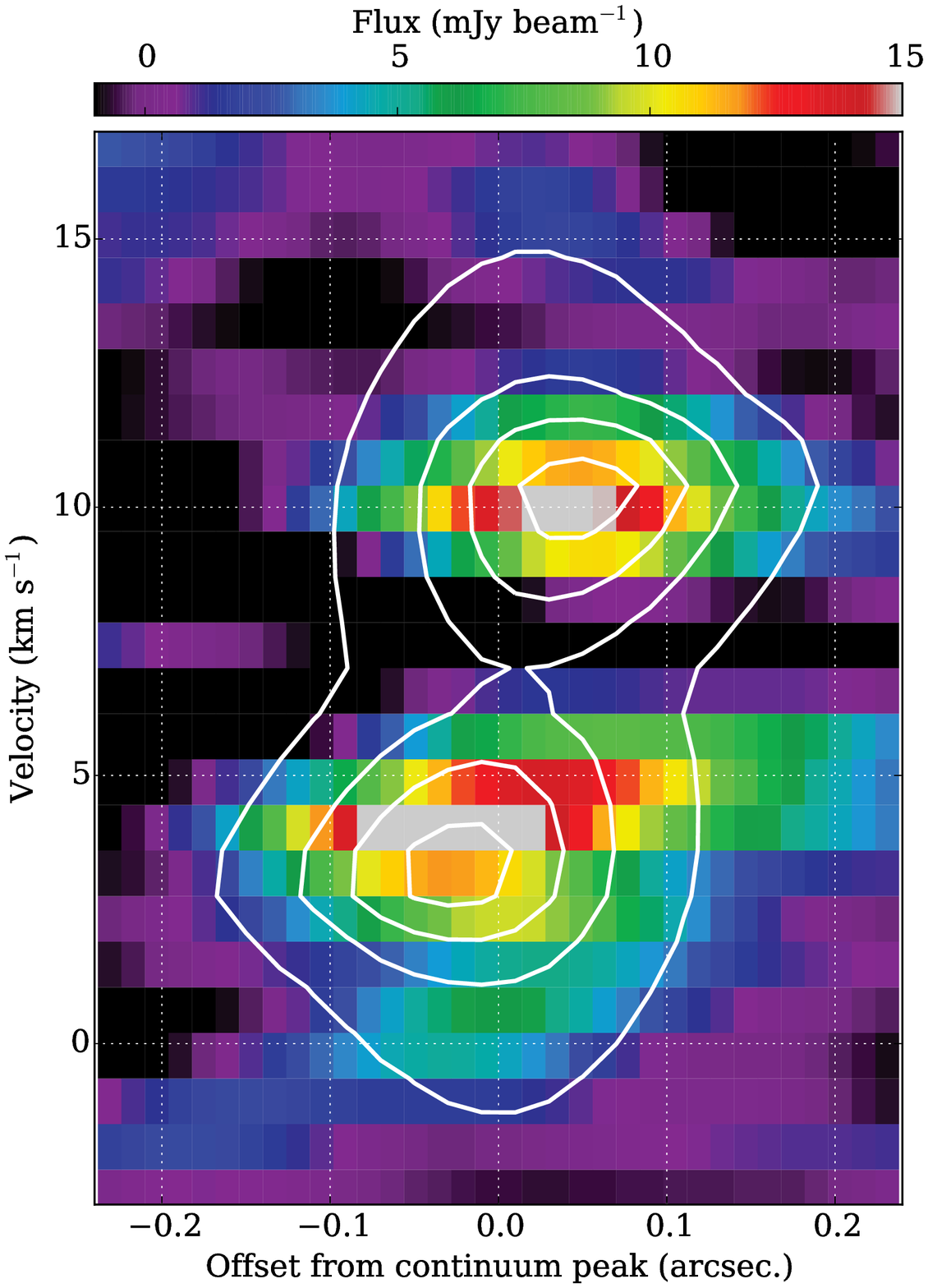}
\caption{The color-scale image shows $^{12}$CO ($J$ = 3--2) P-V diagram of the ALMA observation along the major axis of the disk. Contours show that of the simulated Keplerian model with $M_{*} = 0.18\,M_{\odot}$ (see text). The lowest contour level and subsequent step are 1.0 mJy\,beam$^{-1}$ and 3.0 mJy\,beam$^{-1}$, respectively.
\label{fig:PV}}\end{figure}

In this section, we estimate the dynamical mass of the central protostar, $M_{*}$, and the physical properties of the disk by comparing the observed $^{12}$CO spatial/velocity distributions with those of simulated models. The observed highest velocity toward MMS-1 is $\sim$7 km\,s$^{-1}$ with respect to the systemic velocity at $\sim$2 au from the position of the 0.87 mm continuum peak (Figure \ref{fig:plot}), and it is comparable to the dynamical velocity (e.g., Kepler velocity) for a $M_{*}$ of $\sim$0.1 $M_{\odot}$ (= [$v^2$$r$/$G$]), indicating that there is a sub-solar mass protostar in this source. However, since the present ALMA observations do not have a sufficient resolution to spatially resolve the disk structure, the comparisons with the simulation of rotating disks are needed to remove the effect of the beam dilution. \\
\ In the present study, we assume a Keplerian rotation, as the simplest modeling approach, to estimate the mass of the protostar and the disk properties.
The temperature and density profiles of the simulated models are assumed to be $\propto$ $r^{-0.5}$ and $\propto$ $r^{-1}$. Expected line emissions along lines of sight were calculated with radiative transfer equations, assuming the local thermal equilibrium \citep[see, e.g.,][]{Ohashi14}, and then we simulated the ALMA observations in $^{12}$CO ($J$ = 3--2) and 0.87 mm continuum with the C40-6 configuration, which is the same as in our present observations, using tasks $``$simobserve$"$ and $``$simanalyze$"$ in CASA.  The free parameters of the model are the central stellar mass ($M_{*}$), the disk mass ($M_{\rm disk}$), the disk radius ($R_{\rm disk}$), the position angle (PA), the inclination angle ($i$), the kinematic temperature at a radius of 1 au ($T_{\rm 1au}$)  and systemic velocity ($V_{\rm sys}$). 
We adopted the $V_{\rm sys}$ = 6.75 km\,s$^{-1}$ because the high-velocity components are symmetric around this velocity (see Figure \ref{fig:MMS-1} (b) for the mirrored spectrum).  This velocity is also consistent with the systemic velocity of the dense core observed with the single-dish observations \citep[e.g.,][]{Onishi99}.  We also adopted the inclination angle = 70$\arcdeg$ derived from the scattered light seen in the $Spitzer$ \citep{Terebey09}, and the position angle = $-$17$\arcdeg$ as described in Sec. \ref{subsec:MMS-1_obs}.
Thus, the remaining free parameters are the $M_{*}$, $R_{\rm disk}$, $M_{\rm disk}$, and $T_{\rm 1au}$. We tuned the parameters to reproduce the $^{12}$CO profile at the high-velocity parts because the spectrum is affected by the surrounding dense core toward the source.  We derived the optimum values with the following three steps. (1)
The peak positions of the $^{12}$CO ($J$ = 3--2) in each channel (Figure \ref{fig:chanmap}) with both the ALMA observations and the synthetic observations of Keplerian models are derived by 2D Gaussian fitting.  Based on the plots of the relative velocities to the systemic velocity with respect to the projected distances from the 0.87 mm continuum peak in MMS-1, the parameters of the model were adjusted to match the observations (Figure \ref{fig:plot}). Similar data analysis was carried out in \cite{Tobin12}. The braking of the power law around $\sim$5 au in Figure \ref{fig:plot} is due to the superposition of rotation velocities projected to our line of sight at large radii (see also \citealt{Tobin12}). 
$M_*$ is basically constrained by the highest velocity and $R_{\rm disk}$ is by the breaking distance. These parameters are insensitive to $M_{\rm disk}$ and $T_{\rm 1au}$.  We initially assumed $M_{\rm disk}$ of 8 $\times$ 10$^{-5}$ $M_{\odot}$ and $T_{\rm 1au}$ of 200 K for the simulation, and iterate the simulation after steps (2) and (3).  
(2) We then fitted the $T_{\rm 1au}$ to reproduce the intensity of the spectrum at high-velocity parts.  $M_{\rm disk}$ does not affect the intensity, because the disk is optically thick in $^{12}$CO.  (3) The remaining parameter, $M_{\rm disk}$, was calculated to reproduce the observed 0.87 mm  continuum flux by assuming the temperature distribution derived in step (2). 
The derived optimum values of the Keplerian disk model are $M_{*}$ = 0.18 $M_{\odot}$, $R_{\rm disk}$ = 9 AU, $M_{\rm disk}$ = 8 $\times$ 10$^{-5}$ $M_{\odot}$ and $T_{\rm 1au}$ = 180 K. Our observation and model agree reasonably well (see also the channel maps in Figure \ref{fig:chanmap} and the P-V diagram in Figure \ref{fig:PV}).  The uncertainty of the $M_{*}$ in this analysis is roughly $\pm$0.05 $M_{\odot}$. The inner slopes of models with the $M_{*}$ = 0.13, 0.23 $M_{\odot}$, which have the same other parameters as shown above, are also plotted in Figure 3 for the comparison.\\
\ We note that we simulated only the disk component, although the source is heavily embedded in a dense core.  This difference is seen in the spectrum mainly around the systemic velocity (4.45--9.55 km\,s$^{-1}$); the dense core is observed in this velocity range.  In the redshifted ($>$12.10 km\,s$^{-1}$) range, $^{12}$CO ($J$ = 3--2) observations in Paper II also failed to detect a redshifted outflow seen in the HCO$^+$ ($J$ = 3--2) (Paper I) in the same velocity range due to the absorption by the foreground gas component of the dense core (see also Sec. 3.3 in Paper II).  It is likely that the redshifted ($>$12.10 km\,s$^{-1}$) gas components of the disk are not seen for the same reason.  Our simulation also assumed the symmetric gas distribution of the disk.  There is a slight excess in the observed profile against the model at $-$0.65 km\,s$^{-1}$ (Figure \ref{fig:MMS-1} (b)).  This may be due to the asymmetry of the  gas distribution or to the existence of the velocity component in the same line of sight.  The detailed disk properties is a subject of the future high-angular resolution observations.

\section{Discussions} \label{sec:discussions}

\subsection{A detached disk at the center of an isolated protostellar core \label{subsec:detached}}
Recent interferometric observations have been reporting diversities of Keplerian disks around Class 0/I protostars. Some sources have the disk radii of 50--350 au with the masses of 10$^{-3}$--10$^{-1}$ $M_{\odot}$ \citep[e.g.,][]{Tobin12,Ohashi14,Aso17,Yen17}, however, others have no Keplerian disks \citep[e.g.,][]{Yen15}. While MHD simulations of disk formation \citep[e.g.,][]{Tomida17} tend to predict heavy disks, young circumstellar disks have various masses and sizes. The natures of disk at early stage of star formation are not fully understood so far.\\
\ The central stellar mass of $\sim$0.2 $M_{\odot}$ in MMS-1 is not much different from other Class 0/I objects, and, on the other hand, the disk mass is more than an order of magnitude smaller than those of the Keplerian disks around them. Such small disk mass is normally seen around T Tauri stars \citep{Andrews05}. Because MMS-1 is indeed embedded at the center of the centrally concentrated dense core, it is considered to be still at an early stage of star formation rather than evolved objects. 
The presence of the small disk seen in MMS-1 may be a key to understand the above mentioned diversities of the Keplerian disks around the Class 0/I sources. \\
\ The low luminosity of MMS-1 indicates a low mass accretion rate.   
Even if the luminosity of MMS-1 is dominated by the accretion luminosity, the mass accretion rate is estimated to be quite low as $<$ 2 $\times$ 10$^{-8}$ $M_{\odot}$\,yr$^{-1}$ with the assumption of a stellar radius of 1.5 $R_{\odot}$.
This is consistent with the fact that the outflow activity is quite low (Paper I,II), although a much higher mass accretion rate is needed to form a $\sim$0.2 $M_{\odot}$ protostar within a time scale of the dense core, $\sim$4 $\times$ 10$^{5}$ years  \citep[e.g.,][]{Onishi02}. The existence of the large bipolar nebula with the size scale of $>$1000 au seen in the $Spitzer$ observations \citep{Bourke06,Terebey09} may be evidence of such large accretion activities in the past. In Paper II, we argued the observed density profile of MC27/L1521F, which follows $n(r)$ $\propto$ $r^{-1.4}$ in the inner 3000 au and $\propto$ $r^{-2.0}$ outward of 3000 au,
cannot be explained by the inside-out collapse model \citep{Shu77}, as the central protostar was considered to be at a very young evolutionally stage with an age of $\lesssim$10$^4$ years and the mass of $<$0.1 $M_{\odot}$. We reconsider here the protostar formation with the mass of $\sim$0.2  $M_{\odot}$ in this system. 
In the inside-out collapse model, the expansion wave propagates outward with a sound speed of 0.2 km\,s$^{-1}$ at 10 K.  The time scale of the expansion wave reaching the breaking radius of 3000 au becomes 7 $\times$10$^4$ years.  If we assume the accretion rate after the protostar core formation to be 1.6 $\times$ 10$^{-6}$ $M_{\odot}$\,yr$^{-1}$ \citep{Shu77}, the mass of the formed star becomes $\sim$0.1 $M_{\odot}$ within the time scale; the inside-out collapse model can explain the observed density profile in this case.  If we assume the higher accretion rate to be 1 $\times$ 10$^{-5}$ $M_{\odot}$\,yr$^{-1}$ \citep{Larson69,Hunter77}, the time scale becomes much shorter to achieve the present observed stellar mass, and the expansion wave cannot reach the slope breaking radius of 3000 au.\\
\ Although we cannot clearly describe the mechanisms to create the detached disk without the significant mass accretion at an early stage of star formation so far, a highly dynamical (turbulent) environment seen in this system (Paper I, II) may be a hint to understand the formation of the disk.
Several truncated disks with a size scale of $\sim$10 au toward young brown dwarfs \citep{Ricci12,Testi16} have been also reported with ALMA. \cite{Testi16} suggested that such truncated disks created by dynamical interactions \citep{Bate09,Bate12}. They also confirmed the disk sizes in $\rho$Oph, a cluster-forming region, are much smaller than those in Taurus, which is a relatively quiescent star-forming region, due to their different environments. Our present target, MC27/L1521F, is also located in Taurus, however, the central region shows highly complex spatial/velocity structures compared to other isolated protostellar cores \cite[e.g.,][]{Ohashi14,Evans15,Yen15}, indicating that it resides in the dynamical environment (Paper I,II). There is a possibility that the intrinsically larger disk was stripped by the surrounding gas in the turbulent environment (Paper I). The present $^{12}$CO ($J$ = 3--2) observations also found very high-temperature ($\sim$50 K) gas component at several hundred au away from MMS-1, indicating that there was shock heating possibly caused by interactions among the turbulent gases. The detailed results will be presented in a forthcoming paper (K. Tokuda et al. in preparation).\\
\ We discuss future evolution of MMS-1. Recent numerical simulations that follow evolution of an isolated dense core to a protostar predicted that the mass accretion onto the protostar takes place through the surrounding disk structures \citep[e.g.,][]{Machida08}. The observed disk in MMS-1 is very small, indicating that further significant mass accretion through the disk does not occur anymore. Thus, we may be witnessing a stage close to the moment when the accretion activity has been halting.

\subsection{A very low-luminosity object as a candidate evolved by cold accretion \label{subsec:coldaccretion}}
In this section, we discuss how the low luminosity of $<$0.1 $L_{\odot}$ is possible for a $\sim$0.2 $M_{\odot}$ protostar.  According to the classical evolutionary tracks of pre-main-sequence stars \citep[e.g.,][]{Baraffe98}, it takes $>$1 Myr for a $\sim$0.2 $M_{\odot}$ protostar to reach a luminosity of $<$0.1 $L_{\odot}$. The timescale is much longer than that expected by the present observations as discussed in Sec. \ref{subsec:detached}.  The possible scenario to overcome this issue is thus either the accretion mechanism is different from that or the intrinsic luminosity is underestimated.  The uncertainty of the luminosity may come from the incomplete modeling in an environment of a high ISRF luminosity ($L_{\rm bol}$ = 0.36 $L_{\odot}$) and/or an anisotropic radiation from the optically thick disk, although \cite{Bourke06} and \cite{Terebey09} claim that it is unlikely that the intrinsic luminosity exceeds 0.07 $L_{\odot}$.  The stellar luminosity can be very low even just after the accretion ceases when the accreting gas provides very low entropy (cold accretion;  \citealt{Hartmann97,Hosokawa11,Baraffe12,Vorobyov17,Kunitomo17} and references therein).
We calculated the evolution of the stellar luminosity with the cold accretion by using the numerical code described by \cite{Hosokawa09} and \cite{Hosokawa10,Hosokawa11}. We used a fixed accretion rate of 10$^{-5}$ $M_{\odot}$\,yr$^{-1}$ until the star has reached the mass of 0.2 $M_{\odot}$, and then we stopped the accretion and tracked the evolution of pre-main-sequence star. 
For the cold accretion case, the luminosity depends on the initial stellar radius.
Our calculation starting with a 0.01 $M_{\odot}$ star with the initial radius of 0.65 $R_{\odot}$ reproduces the observed luminosity ($\sim$0.07 $L_{\odot}$) in $<$10$^5$ years after we stop the mass accretion. In this case, we note that much smaller initial stellar radius than that calculated by \cite{Masunaga00}, 4 $R_{\odot}$, is required to produce the low luminosity.
As described above, it is possible, in principle, to make the luminosity of a very young 0.2 $M_{\odot}$ protostar less than 0.1 $L_{\odot}$ just after the end of mass accretion if we adopt the cold accretion model. Mechanisms to create such small stellar core and environments that enable the cold accretion are currently unknown. Although more observations toward similar objects are required to obtain a conclusive interpretation of these objects, the present observation provides a challenge to the current theory for the formation and early evolution of a protostar. 

\section{Conclusions}
ALMA observations of a very low-luminosity object (VeLLO) in a dense core, MC27/L1521F, have obtained strong evidence of a rotating disk with the central stellar mass of $\sim$0.2 $M_{\odot}$, the disk mass of $\sim$10$^{-4}$ $M_{\odot}$, and the disk radius of $\sim$10 au with a quite low accretion activity as a protostar. We suggest that the compact detached disk might be created by the interaction with the surrounding gas in a turbulent environment in this system. The current low mass accretion rate may indicate that the mass accretion onto the protostar/disk has halted quite recently. 
Our present results also indicate the low-luminosity ($<$0.1 $L_{\odot}$) of sub-solar-mass protostar can be explained by cold accretion model. These facts indicate that the gas dynamics at early stage of star formation largely affects the mass accretion history. 

\acknowledgments
This paper makes use of the following ALMA data: ADS/ JAO.ALMA\#2015.1.00340.S. ALMA is a partnership of the ESO, NSF, NINS, NRC, NSC, and ASIAA. The Joint ALMA Observatory is operated by the ESO, AUI/NRAO, and NAOJ. This work was supported by NAOJ ALMA Scientific Research Grant Numbers 2016-03B and JSPS KAKENHI (Grant No. 22244014, 23403001, and 26247026).

\end{document}